\documentstyle[emulateapj]{article}
\def\hi{\sc H$\,$i~}
\def\kms{km s$^{-1}$}

\begin{document}

\title{REDSHIFTS OF GALAXIES AROUND ARP 220 AND
       SERENDIPITOUS DISCOVERY OF THREE STAR-FORMING DWARF GALAXIES
       AT REDSHIFT $z \sim$ 0.5}
\author{Youichi Ohyama$^1$, Yoshiaki Taniguchi$^1$, J. E. Hibbard$^2$, \&
William D. Vacca$^3$}

\vspace {1cm}

\affil{$^1$Astronomical Institute, Tohoku University, Aoba, Sendai 980-8578,
Japan}
\affil{$^2$National Radio Astronomy Observatory, 520 Edgemont Rd.,
Charlottesville,
           VA 22903, U.S.A.}
\affil{$^3$Institute for Astronomy, University of Hawaii, 2680 Woodlawn Drive,
           Honolulu, HI 96822, U.S.A.}


\begin{abstract}
We present redshift measurements of four faint galaxies around the
archetypal ultraluminous infrared galaxy Arp 220. These galaxies
have significantly higher redshifts ($z \sim$ 0.036 -- 0.091) than
that of Arp 220 ($z \simeq 0.018$).  Therefore, we conclude
that they are background objects not physically related to Arp 220.
Three of these faint galaxies located to the south of Arp 220 are a group of
galaxies (or the brightest members in a cluster of galaxies) at $z
\simeq 0.09$, as suggested by Heckman et al. [1996, ApJ, 457, 616] on
the basis of their associated soft X-ray emission.

We also report the serendipitous discovery of three additional
galaxies at redshift $z \sim 0.5$, found along one of the slit
positions.  All three galaxies exhibit an [O {\sc ii}]$\lambda$3727
emission line.  The spectrum of the brightest galaxy ($m_R \simeq
24.4$) shows other strong emission lines: Mg {\sc ii} $\lambda$2798,
H$\beta$, [O {\sc iii}]$\lambda$4959, and [O {\sc iii}]$\lambda$5007.
The emission-line properties of these galaxies as well as their
intrinsically low luminosities ($M_R \geq -18.4$) indicate that they
are star-forming dwarf galaxies.
\end{abstract}



\section{INTRODUCTION}

Arp 220 (IC 4553) is the archetypal ultraluminous infrared galaxy
($L_{\rm IR} > 10^{12} L_\odot$; Soifer et al. 1984), and has been 
discussed extensively in the literature, mostly with regard to 
its huge infrared luminosity (Emerson et al.~1984; Heckman, Armus 
\& Miley 1987; Sanders et al. 1988; Shaya et al. 1994; Genzel et 
al.  1998; Taniguchi, Trentham, \& Shioya 1998; Taniguchi \& Ohyama
1998; for a review see Sanders \& Mirabel 1996).  Recently, Heckman et
al. (1996) presented their ROSAT observations of the soft X-ray
properties of Arp 220.  Interestingly, they find an additional diffuse
soft X-ray source lying $\sim 2^\prime$ to the south-southwest, with
an apparent bridge connecting back to Arp 220.  Since Arp 220 is also
a well-known superwind galaxy (Heckman et al. 1987, 1990; Taniguchi et
al. 1998), this extension could be interpreted as a result of a
wind-blown ``supperbubble''. However, this X-ray source is located in
a direction approximately perpendicular to the proposed outflow axis
for the wind (PA$\approx 135^\circ$ from HRI imaging of Heckman et
al.~1996), and Heckman et al.~suggest instead that it is associated
with a small compact group of galaxies coincident with the emission
(see Figure 1). They note that the soft X-ray properties of the
southern source (size, luminosity, and spatial extension) can be
explained if this group is located at redshift $z \sim 0.1$.

However, recent VLA observations of the neutral hydrogen in Arp 220 by
Yun, Hibbard \& Scoville (1999; see also Hibbard \& Yun 1996, 1999)
call into question the interpretation given by Heckman et al
(1996). These observations show that the {\hi} gas directly associated
with Arp 220 has an anomalous extension that projects directly onto
the same group of galaxies. This raises the possibility that one or
more of the galaxies in the group is a gas rich-rich system associated
with Arp 220 rather than a background object.  We therefore targeted
the group of galaxies for spectroscopic observations to directly
determine their redshifts.

In order to directly determine the redshifts of the group, we acquired 
optical spectra of these galaxies with the Keck II telescope. This paper
reports on the result of these spectroscopic observations,
which include the three galaxies to the south-southwest as well as a
larger face-on barred galaxy located to the southeast of Arp 220.  The
optical identifications of the four galaxies are shown in Figure 1.
The three galaxies labeled as A, B, and C are the putative background
group identified by Heckman et al. (1996). The galaxy labeled as D is
IC 4554, for which there are no photometric or spectroscopic
observations available in the literature.

We have serendipitously discovered three star-forming galaxies at $z\sim 0.5$
during these observations.
The analysis of their spectra will be reported in section 3.3.

We adopt a Hubble constant $H_0$ = 50 km s$^{-1}$ Mpc$^{-1}$ and a
deceleration parameter $q_0 = 0.5$ throughout this paper.


\section{OBSERVATIONS}

\subsection{Optical Spectroscopy}

Optical spectra of the visual companions to Arp 220 were acquired with
the Low Resolution Imaging Spectrograph (LRIS; Oke et al. 1995) on the Keck II
telescope on the nights of 1997 April 15 and 16 (UT) during a break in
another observing project.  On the first night (15 April) a 900 s
exposure was obtained with a $1''$ slit centered on galaxy A (see
Figure 1) and rotated to a position angle of 156 degrees. The 300 l/mm
grating used for these observations is blazed at 5000 \AA\ and yields
a dispersion of 2.55 \AA /pixel. (The scale in the spatial direction
is $0.215''$/pixel.)  The central wavelength was set to 6565
\AA\ and the resulting spectrum covers the wavelength range 3990 --
8970 \AA. The resolution was measured to be 13 \AA. Internal
calibration lamp exposures, for both wavelength calibration and flat
fields, were taken immediately after the science exposure at the same
sky position.  The error of the wavelength calibration is 0.1 \AA~
over the entire wavelength range of the spectra.  Although the slit
was rotated to cover the companions A and C, and was not set at the
parallactic angle, the airmass during the observations was
approximately 1.10, and we therefore made no corrections to the
observed flux to account for atmospheric dispersion.  Spectra of the
standard stars Feige 67 and BD +28$^\circ$ 4211 were obtained with the
slit rotated to the parallactic angles for the purposes of flux
calibration.  Sky conditions were photometric during these
observations.

On the second night, we obtained optical spectra of companion D. For
these observations, the $1''$ slit was again used and the position
angle was set to 130 degrees. A 900 s exposure was taken with the 600
l/mm grating, with a central wavelength setting of 6024 \AA.  The
grating is blazed at 5000 \AA\ and the dispersion is 1.28 \AA /pixel.
The observational set-up yielded wavelength ranges and resolutions of
4670 -- 7210 \AA\ and 6.9 \AA, respectively.  The uncertainty in the
wavelength calibration is 0.05 \AA\ over the entire wavelength range
of the spectra.  The exposure was made at airmass of $\sim 1.2$.
Exposures of internal lamps were obtained for wavelength calibration
and flat-field purposes.  Spectra of standard stars Feige 34, Feige
67, and BD +33$^\circ$ 2642 were obtained with a wide slit ($8.7''$)
rotated to the parallactic angle for flux calibration.  The sky
conditions were photometric during these observations, and the seeing
was estimated to be $\sim 1''$.

Data reduction was performed using IRAF.
The optical spectra of the four galaxies were extracted 
from three rows centered on the observed peak brightness of each
galaxy, corresponding to an aperture of 0.64 $\times$ 1 arcsec$^2$.
The spectra are shown in Figure 2 and the line identifications and the
measured redshifts are given in Table 1.

\subsection{Optical Imaging}

Two 600 sec $R$-band CCD images were obtained with the Tek 2k CCD
camera mounted at the f/10 Cassegrain focus of the University of
Hawaii 2.2 m telescope at Mauna Kea Observatories on 1995 June 3.  The
pixel scale was 0.22\arcsec, which yields a field of view of
7.5\arcmin.  Since the seeing was measured to be 1.2\arcsec, the
pixels were subsequently rebinned $2\times 2$.  The conditions were
photometric on both dates, and the data were calibrated via
observations of Landolt UBVRI standards (Landolt 1983) observed on the
same nights. The zero point errors (1$\sigma$) were measured to be
0.03 mag. The images were flattened and combined using the techniques
described in Hibbard \& van Gorkom (1996), and the final mosaics are
flat to better than one part in 500.

We have performed aperture photometry using FOCAS (Jarvis \& Tyson 1981)
and measured the total magnitudes of the four galaxies.
The total magnitudes are measured in the following manner:
For each galaxy, a contour at the 4$\sigma$ level above the sky
background was determined. We then define a homologous contour whose
surface is twice as large as that of the 4$\sigma$ contour.  The total
magnitude is measured using this larger aperture. 
These magnitudes are corrected for 0.1 mag of Galactic extinction 
in the R band (0.15 mag in the B band, Burstein \& Heiles 1984) using the 
average Galactic extinction law of Seaton (1979). 
The results are summarized in Table 2.
Note that we do not apply K
correction in the estimates of absolute $R$ magnitudes. In Table 2, we
also give positions of the four galaxies measured from the
digitized sky survey (DSS) image.  Although the morphological
classification of galaxy D (IC 4554) given in the NASA
Extragalactic Database (NED) is S0, our CCD image (Figure 1) shows an
evident bar structure in this galaxy.

\section{RESULTS AND DISCUSSION}

\subsection{Redshifts and Relation to Arp 220}

The three galaxies, A, B, and C, have nearly identical redshifts
($z=$0.088---0.091). The galaxy D (IC 4554) has a redshift of 0.036.
These redshifts are much higher than that of Arp 220 ($z = 0.018$) and
thus all four galaxies are background galaxies with no physical
relation to Arp 220 (however see Arp 1987).  As shown by Heckman et
al. (1996), the soft X-ray properties of the southern extension can be
reasonably understood in terms of the expected X-ray emission
associated with a galaxy group or poor cluster.  Accordingly, our
optical spectroscopy has confirmed the interpretation proposed by
Heckman et al. (1996). Since these redshifts are much higher than that
of the {\hi} in this area ($V_{\rm HI}\approx 5550$ \kms; Yun et
al. 1999), we conclude that the apparent {\hi} extension towards the
southern group is a chance projection effect.

If we adopt a redshift $z = 0.091$ for the southern group, the
luminosity distance is estimated to be 558 Mpc.  Thus the absolute $R$
magnitudes of the two brightest galaxies are $M_R \simeq -23$. This
magnitude is comparable to those of first-ranked cluster elliptical
galaxies (Postman \& Lauer 1995).  In addition, the redshift
difference among the three galaxies ($\Delta z\sim 0.003$) corresponds
to a velocity difference of 900 km s$^{-1}$, which is comparable to
the typical velocity dispersion of a cluster of galaxies. Therefore,
we conclude that these galaxies are physically associated and comprise
a group of galaxies or the brightest galaxies in a cluster of galaxies
at $z \simeq 0.09$.

\subsection{Comments on the Emission-line Galaxy B}

The optical spectroscopy reveals that galaxy B is an emission-line 
galaxy. The observed emission line fluxes are listed in Table 3, and 
here we use them to derive the characteristics of the ionized gas and 
the hot star population.  

After correcting for the Galactic extinction ($A_B = 0.15$ mag),
we obtain a Balmer decrement, $F$(H$\alpha$)$/F$(H$\beta$) $\simeq 3.44
\pm 0.09$. This yields a visual extinction $A_V \simeq 0.51 \pm 0.10$ mag
(Case B with the electron temperature $T_{\rm e} = 10^4$ K; Osterbrock
1989). This value is used to obtain the reddening-corrected fluxes
given in Table 3. The [S {\sc ii}] doublet ratio, $I$([S {\sc
ii}]$\lambda$6717)$/I$([S {\sc ii}]$\lambda$6731) $\simeq 1.35 \pm
0.06$, gives an electron density of $n_{\rm e} \simeq 70$
cm$^{-3}$. The [N {\sc ii}]$\lambda$6583/H$\alpha$, [S {\sc
ii}]($\lambda$6717 + $\lambda$6731)/H$\alpha$, and [O {\sc
iii}]$\lambda$5007/H$\beta$ line ratios indicate that massive stars
are the source of the ionizing photons (Veilleux \& Osterbrock 1987).
For a distance of 558 Mpc ($z = 0.091$), the reddening-corrected
H$\beta$ luminosity, $L$(H$\beta$) $\simeq 9.93 \times 10^{39}$ erg
s$^{-1}$, yields a production rate of Lyman continuum photons of
$N$(Lyc) $\simeq 2.08 \times 10^{52}$ photons s$^{-1}$, equal to the
output of $\sim 610$ O5V stars (Vacca, Garmany, \& Shull
1996). Finally, the ratio $R_{23} \equiv$ [$I$([O {\sc
ii}]$\lambda$3727) + $I$([O {\sc iii}]$\lambda$4959) + $I$([O {\sc
iii}]$\lambda$5007)]/$I$(H$\beta$) can be used to estimate the
metallicity of the galaxy.  Using the reddening-corrected fluxes given
in Table 3, we obtain $R_{23} \simeq 4.60 \pm 0.06$.  The empirical
calibration of Edmunds \& Pagel (1984, see also Vacca \& Conti 1992)
then yields 12 + log (O/H) $\simeq 8.64$, or an oxygen abundance with
respect to solar of [O/H] $\simeq -0.29$. (We adopted the solar oxygen
abundance value (O/H)$_\odot$ = 8.303$\times 10^{-4}$ of Meyer 1985.)
The derived properties of galaxy B are typical of those of
star-forming galaxies.

\subsection{Serendipitous Discovery of Three Star-Forming Dwarf
Galaxies at $z \sim 0.5$}

In the two-dimensional spectrum taken at PA $156^\circ$ on 1997 April
15, we serendipitously found continuum and line emission from three
additional faint galaxies. Their spectra are shown in Figure 3, where
the galaxies have been denoted as I, II, and III (brightest to
faintest). Continuum emission from another very faint galaxy (denoted
as IV) is also seen in Figure 3. The spectra of galaxies I, II, and
III were extracted from three rows (0.64\arcsec) centered on the
observed peak brightness of each galaxy.  In order to identify these
galaxies, we show a close up of our $R$-band CCD image in Figure 4.
The faintest galaxy IV in Figure 3 appears to be as bright as galaxy I
in this image. Thus the faintness in Figure 3 may be due to our slit
covering only a portion of this galaxy.  Since there are no convincing
emission lines in this spectrum, we will not discuss this galaxy
further in this paper.

The coordinates of galaxies I, II, and III
are given in Table 4. 
We also give rough estimates of $R$ magnitudes which
were measured using the spectra shown in Figure 3.
The observed magnitudes were corrected for 0.1 mag of Galactic extinction
in the R band (0.15 mag in the B band, Burstein \& Heiles 1984) using the
average Galactic extinction law of Seaton (1979).
Note that we do not apply K
correction in the estimates of absolute $R$ magnitudes.
Since it is unlikely that our slit covered the entire images of the galaxies,
the magnitudes given in Table 4 
may be upper limits.

 Identifications, fluxes, and equivalent widths of the emission lines,
 as well as the redshifts derived from the observed wavelengths of the
 lines, are summarized in table 5.  All the galaxies are located at $z
 \simeq 0.5$.  Note that galaxy III has only one obvious emission line
 which we identify as [O {\sc ii}]$\lambda$3727. We see a faint
 emission feature which may be H$\beta$ in Figure 3 but we cannot
 confirm it unambiguously. The widths of the emission lines are not 
resolved by our spectroscopy.  Our resolution (13 \AA) yields an upper 
limit to the line widths of $\simeq$ 700 km s$^{-1}$ for the redshifted 
[O {\sc ii}] $\lambda$3727 line. 

Since galaxy I shows several strong emission 
 lines, we investigate its ionization properties in more detail.
 In Figure 5, we show the location of galaxy I in the 
[O {\sc ii}]$\lambda$3727/[O {\sc
 iii}]$\lambda$5007 versus [O {\sc iii}]$\lambda$5007/H$\beta$
 excitation diagram (Baldwin, Phillips, \& Terlevich 1981). For
 comparison, we plot a solid curve representing a
 sequence for H {\sc ii} regions photoionized by massive stars, as well
 the data for nearby irregular and spiral galaxies taken
 from Kennicutt (1992). Since galaxy I lies in the same area of the
excitation diagram as these data, its 
excitation properties are typical of local star forming galaxies.  
 However, it should be noted
 that we have made no correction for the (unknown) internal extinction,
 which could signifcantly change the [O {\sc ii}]$\lambda$3727] line
 flux. 
We also see no evidence for a broad
component at the base of the H $\beta$ line, but our S/N is not
sufficient to allow a definitive statement regarding the presence or
absence of such a feature.

 The spectrum of galaxy I exhibits a very blue continuum as well as strong 
 [Ne {\sc iii}]$\lambda$3869 and Mg {\sc ii} $\lambda$2798 emission lines. 
 The intensities relative to that of 
H$\beta$ are 0.51 and 1.53, respectively (with no reddening corrections).
 The [Ne {\sc iii}]$\lambda$3869 emission line is contaminated by a strong 
 night sky line (see Figure 3).  Although the [Ne {\sc iii}]$\lambda$3869 
 emission line has been detected in a number of star-forming galaxies, 
 Mg {\sc ii} $\lambda$2798 in emission is more typical of AGN (e.g., 
Storchi-Bergmann, Kinney, 
 \& Challis 1995). However, these emission lines may
 be generated in shock-heated regions (Dopita \& Sutherland 1996),
 and it is possible that the ionized gas in galaxy I is affected by some
 wind activity by massive stars and their descendents (e.g., Wolf-Rayet
 stars and supernovae). Furthermore, although Mg {\sc ii} $\lambda$2798 
 emission is generally rare in starburst systems, some starbursts do 
in fact exhibit this emission feature
(e.g. the blue compact dwarf galaxies NGC 1510 and Tol
1924-416; Kinney et al. 1993; Storchi-Bergmann, Kinney, \& Challis
1995). While we have not detected any high ionization lines such as
[Ne {\sc v}], nor see any evidence for a broad line component in the
detected lines, our observations still cannot rule out an AGN origin.
Nonetheless, we will proceed under the assumption that the emission 
lines are due to a starburst.

The emission line luminosities can be used to estimate the star
formation rate (SFR) in these galaxies. We use the calibration based
on [O {\sc ii}] luminosity derived by Kennicutt (1992; SFR $\simeq 5
\times 10^{-41}$ {\it L}([O {\sc ii}]) $M_\odot$ y$^{-1}$).
The results are summarized in Table 6. 
We obtain SFR $\simeq 4 M_\odot$ y$^{-1}$ for galaxy I while
SFR $\simeq 2 M_\odot$ y$^{-1}$ for galaxies II and III.
These SFRs are smaller by a factor of a few than that for a normal galaxy
(i.e., a so-called $L^*$ galaxy), $\simeq 10 M_\odot$ y$^{-1}$
(e.g., Cowie, Hu, \& Songaila 1995).

The estimated parameters for galaxies I, II, and III indicate that they 
are so-called faint blue galaxies, similar to those found in the previous
deep survey programs (e.g., Cowie, Songaila, \& Hu 1991; see for reviews,
Koo \& Kron 1992; Ellis 1997). They may be also related to the class of
compact narrow emission line galaxies at intermediate redshifts
(Koo et al. 1995; Guzm\'an et al. 1996).

The observed separations between the galaxies is 8.4\arcsec,
corresponding to a linear separation $\simeq$ 60 kpc. This is typical
of the mean separations found in compact groups of galaxies (Hickson et
al.~1992).  Such groups frequently contain multiple star forming members
(e.g. Coziol et al.~1998). However, since light from galaxy C prohibits
a clear view of the region around these systems (Fig~4), it is not
possible to say whether we have discovered an intermediate redshift
compact group, or a subcomponent of a larger structure.

\vspace {0.5cm}

We would like to thank the staff of W. M. Keck Observatory for their
kind support for our observations.
We would also like to thank the referee for useful comments.
YO is a JSPS Research Fellow. 
WDV acknowledges partial support in the form of a fellowship from the
Beatrice Watson Parrent Foundation.
JEH thanks D. M. Whittle and T. X. Thuan for useful conversations.
This work was partly supported by
the Japanese Ministry of Education, Science, and Culture
(Nos. 10044052, and 10304013). 
Some of the data presented herein were obtained at the W.M. Keck
Observatory, which is operated as a scientific partnership among the
California Institute of Technology, the University of California and
the National Aeronautics and Space Administration.  The Observatory was
made possible by the generous financial support of the W.M. Keck
Foundation.
This research has made extensive
use of the NASA/IPAC Extragalactic Database (NED), which is operated
by the Jet Propulsion Laboratory, California Institute of Technology,
under contract with the National Aeronautics and Space Administration.
The Digitized Sky Surveys were produced at the Space Telescope Science
Institute under U.S.~Government grant NAG W-2166.  The images of these
surveys are based on photographic data obtained using the Oschin
Schmidt Telescope on Palomar Mountain and the UK Schmidt Telescope.
The plates were processed into the present compressed digital form
with the permission of these institutions.



\newpage

\begin{figure*}
\epsscale{1.0}
\plotone{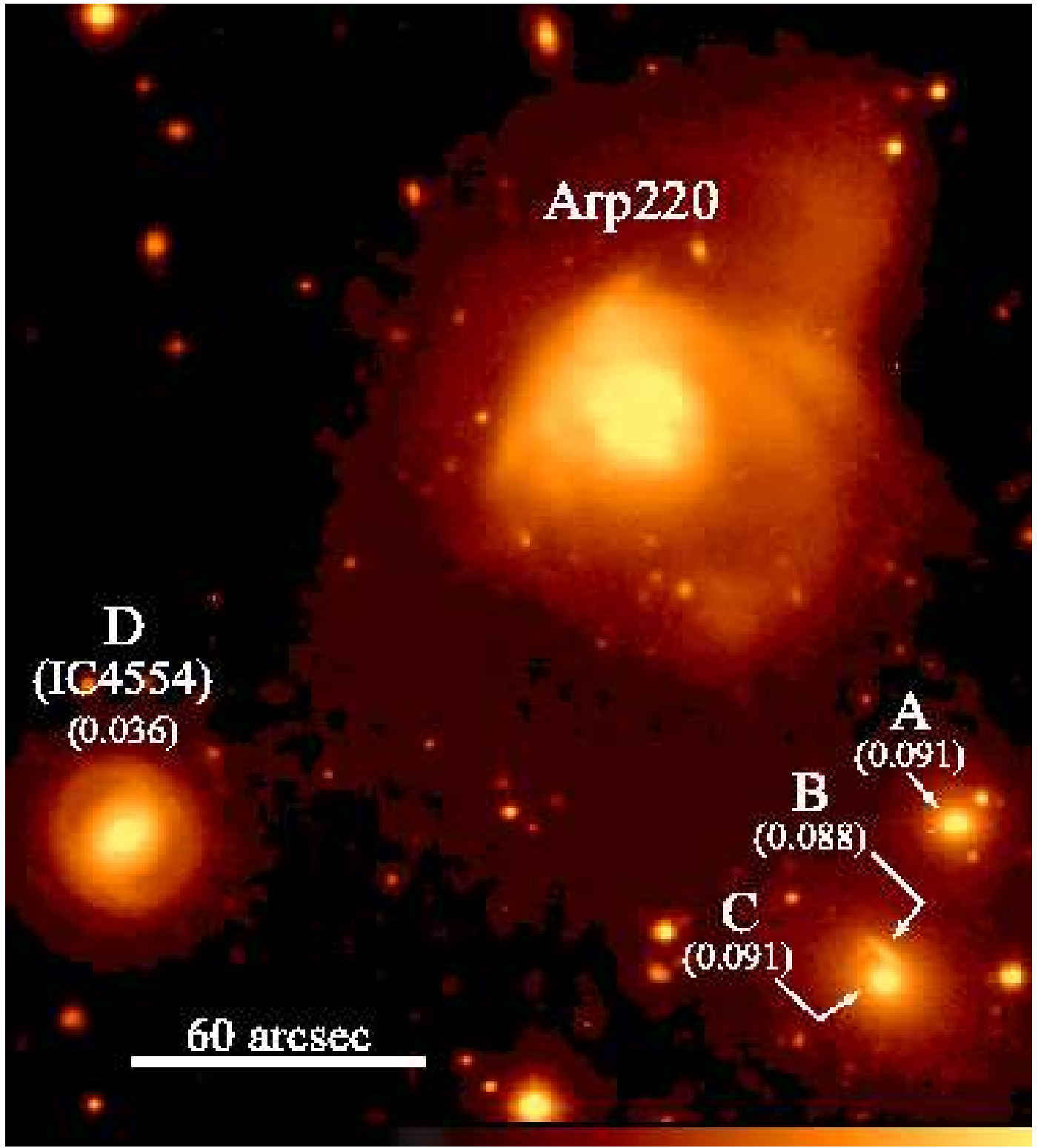}
\caption{Optical identification of the four galaxies around Arp 220.
\label{fig1}}
\end{figure*}

\begin{figure*}
\plotone{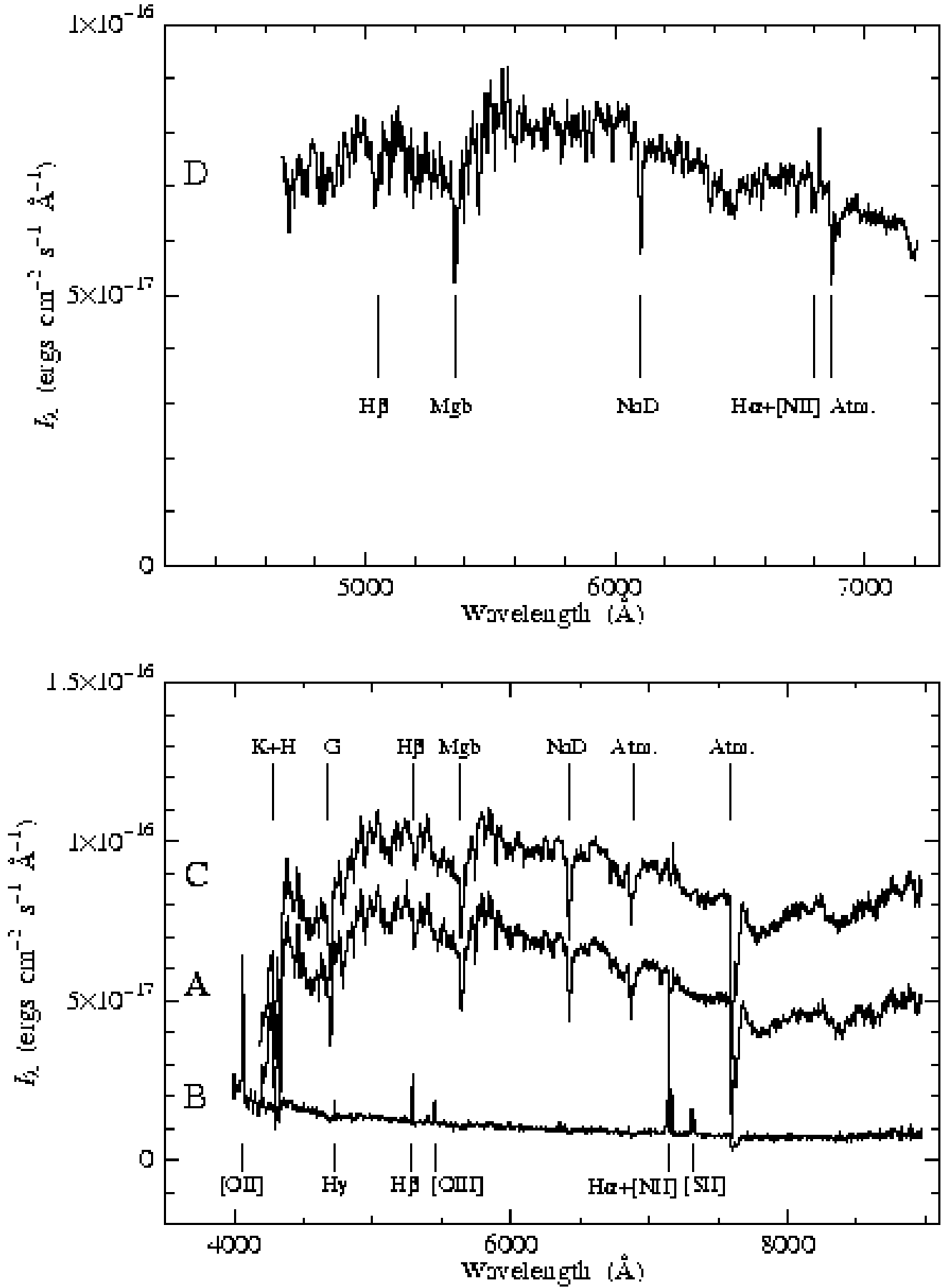}
\caption{Keck spectra of the four galaxies. a) The spectrum of galaxy D,
and b) the spectra of the three galaxies located to the south
of Arp 220 (A, B, and C).
To show the spectra clearly, the zero point of the flux scale of galaxy A
is shifted to $-2\times 10^{-17}$ ergs cm$^{-2}$ s$^{-1}$ \AA$^{-1}$.
\label{fig2}}
\end{figure*}

\begin{figure*}
\plotone{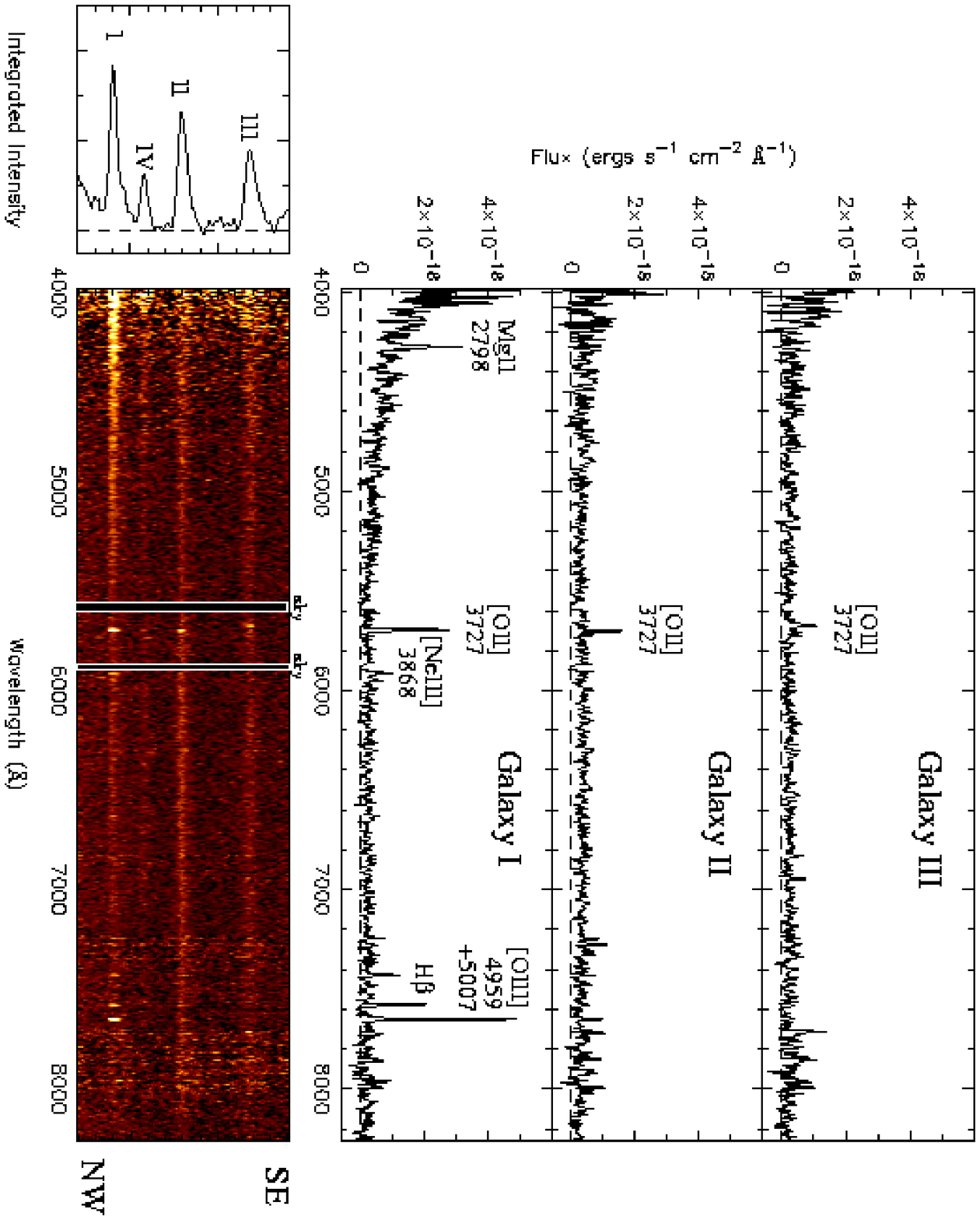}
\caption{Keck spectra of the three star-forming dwarf galaxies.
\label{fig3}}
\end{figure*}

\begin{figure*}
\plotone{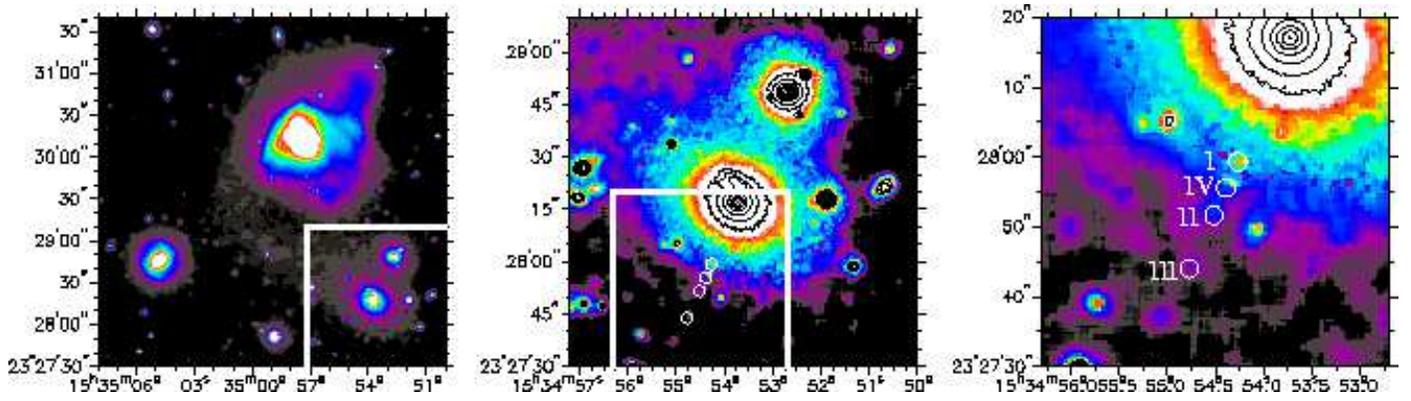}
\caption{Optical positions of  the three intermediate-redshift galaxies
discovered in this study are shown by open circles.
This image is made using the same $R$-band CCD image shown in Figure 1.
\label{fig4}}
\end{figure*}

\begin{figure*}
\epsscale{0.9}
\plotone{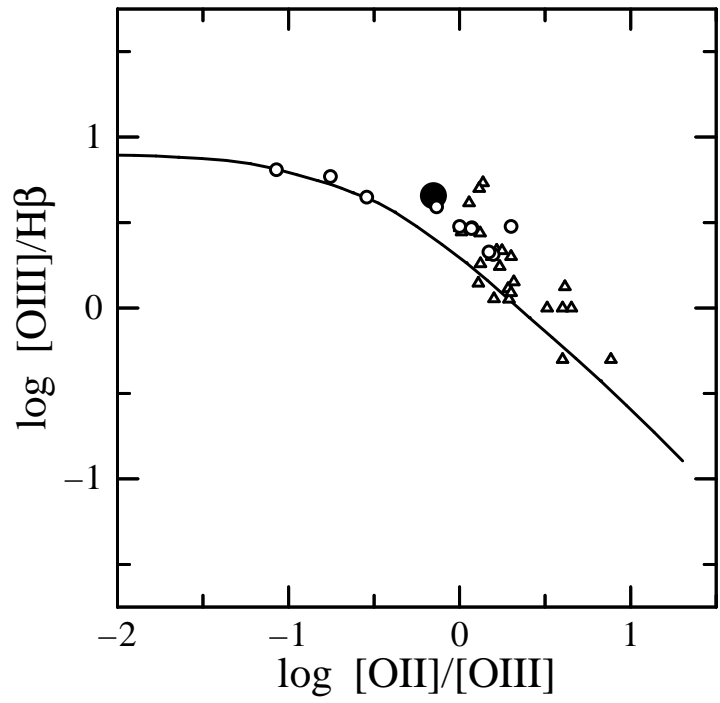}
\caption{
Excitation diagram between
[O {\sc ii}]$\lambda$3727/[O {\sc iii}]$\lambda$5007 and
[O {\sc iii}]$\lambda$5007/H$\beta$. Galaxy I is shown by the large filled
circle. Irregular galaxies and spiral galaxies observed by Kennicutt (1992)
are shown by open triangles and open circles, respectively.
The solid curve shows a sequence for H {\sc ii} regions derived
by Baldwin et al. (1981).
\label{fig5}}
\end{figure*}

\newpage


\begin{deluxetable}{lcccc}
\tablecaption{Redshift of galaxies around Arp220}
\tablehead{
\colhead{Line} &
\colhead{$\lambda_{\rm rest}$ ~ (\AA)} &
\colhead{Emi. or Abs.} &
\colhead{$\lambda_{\rm obs}$ ~ (\AA)} &
\colhead{$z$} \\
}
\startdata
Galaxy A  &        &      &         &        \nl
\tableline
Ca {\sc ii}  K    & 3933.7 & Abs. & 4292.1  & 0.0911 \nl
Ca {\sc ii}  H    & 3968.5 & Abs. & 4329.1  & 0.0909 \nl
CH G band & 4300.0 & Abs. & 4694.5  & 0.0917\tablenotemark{b} \nl
H$\beta$  & 4861.3 & Abs. & 5305.5  & 0.0914\tablenotemark{b} \nl
Mg b      & 5172.7 & Abs. & 5643.7  & 0.0911\tablenotemark{b} \nl
Na D      & 5893.0\tablenotemark{a} & Abs. & 6425.5  &
0.0904\tablenotemark{b} \nl
\tableline
 &  &  & & $<z> = 0.0910\pm 0.0001$ \nl
\tableline
Galaxy B  &        &      &         &        \nl
\tableline
[O {\sc ii}]  & 3727.4\tablenotemark{a} & Emi. & 4058.9 &
0.0889\tablenotemark{b} \nl
H$\gamma$ & 4340.5 & Emi. &  4725.1  & 0.0886 \nl
H$\beta$  & 4861.3 & Emi. &  5291.0  & 0.0884 \nl
[O {\sc iii}] & 4958.9 & Emi. &  5396.4  & 0.0882 \nl
[O {\sc iii}] & 5006.9 & Emi. &  5448.9  & 0.0883 \nl
[N {\sc ii}] & 6548.1 & Emi. &  7122.3  & 0.0877 \nl
H$\alpha$ & 6562.8 & Emi. &  7138.3  & 0.0877 \nl
[N {\sc ii}] & 6583.4 & Emi. &  7160.9  & 0.0877 \nl
[S {\sc ii}] & 6716.4 & Emi. &  7305.5  & 0.0877 \nl
[S {\sc ii}] & 6730.8 & Emi. &  7320.9  & 0.0877 \nl
\tableline
 &  &  &  & $<z> = 0.0880\pm 0.0004$ \nl
\tableline
Galaxy C  &        &      &         &        \nl
\tableline
Ca {\sc ii} K & 3933.7 & Abs. & 4290.6 & 0.0907 \nl
Ca {\sc ii} H & 3968.5 & Abs. & 4326.7 & 0.0903 \nl
CH G band & 4300.0 & Abs. & 4693.0  & 0.0914\tablenotemark{b} \nl
H$\beta$  & 4861.3 & Abs. & 5305.7  & 0.0914\tablenotemark{b} \nl
Mg b      & 5172.7 & Abs. & 5642.0  & 0.0907\tablenotemark{b} \nl
Na D      & 5893.0\tablenotemark{a} & Abs. & 6423.4  &
0.0900\tablenotemark{b} \nl
\tableline
  &   &  &  & $<z> = 0.0905\pm 0.0002$ \nl
\tableline
Galaxy D  &        &      &         &        \nl
\tableline
H$\beta$  & 4861.3 & Abs. & 5037.7  & 0.0363  \nl
Mg b      & 5172.7 & Abs. & 5355.9  & 0.0354\tablenotemark{b}  \nl
Na D      & 5893.0\tablenotemark{a} & Abs. & 6103.0  &
0.0356\tablenotemark{b}  \nl
[N {\sc ii}]     & 6583.4 & Emi. & 6819.0  & 0.0358  \nl
\tableline
 &  &  &  & $<z> = 0.0360\pm 0.0003$ \nl
\enddata
\tablenotetext{a}{Mean wavelength of the doublet.}
\tablenotetext{b}{The redshift is not used for calculating the mean redshift
                  because of the line blending.}
\end{deluxetable}


\newpage

\begin{deluxetable}{lccccc}
\tablecaption{Positions and photometric properties of galaxies around Arp220}
\tablehead{
\colhead{Object ID} &
\colhead{RA(J2000)} &
\colhead{DEC(J2000)} &
\colhead{Morphology} &
\colhead{$m_R$\tablenotemark{a}} &
\colhead{$M_R$\tablenotemark{a}} \nl
\colhead{} &
\colhead{($^h ~ ^m ~ ^s$)} &
\colhead{($^\circ ~ ^\prime ~ ^{\prime\prime}$)} &
\colhead{} &
\colhead{(mag)} &
\colhead{(mag)} \nl
}
\startdata
A & 15 34 52.7 & 23 28 48 & E & 15.9 & $-$22.8  \nl
B & 15 34 53.6 & 23 28 21 & S & \nodata & \nodata \nl
C & 15 34 53.7 & 23 28 17 & E & 15.2\tablenotemark{b} &
$-$23.6\tablenotemark{b} \nl
D & 15 34 04.8 & 23 28 45 & S0 & 14.6 & $-$22.1 \nl
\enddata
\tablenotetext{a}{Corrected for the Galactic extinction (see text).}
\tablenotetext{b}{A total magnitude of both B and C.}
\end{deluxetable}


\newpage

\begin{deluxetable}{lcccc}
\tablecaption{Emission-line fluxes of the galaxy B}
\tablehead{
\colhead{Line} &
\colhead{$F$\tablenotemark{a} ($10^{-17}$ erg cm$^{-2}$ s$^{-1}$)} &
\colhead{$I$\tablenotemark{b} ($10^{-17}$ erg cm$^{-2}$ s$^{-1}$)} &
\colhead{$EW_{\rm obs}$\tablenotemark{c} (\AA)} &
\colhead{$EW_{\rm rest}$\tablenotemark{d} (\AA)} \\
}
\startdata
[O {\sc ii}]$\lambda$3727  & 49.5 & 117.4 & 28  & 26  \nl
H$\gamma$           & 4.7  & 10.3  & 3.7 & 3.4 \nl
H$\beta$            & 13.2 & 26.6  & 13  & 12  \nl
[O {\sc iii}]$\lambda$4959 & 2.4  & 4.7   & 2.0 & 1.8 \nl
[O {\sc iii}]$\lambda$5007 & 5.7  & 11.1  & 5.3 & 4.9 \nl
[N {\sc ii}]$\lambda$6548  & 4.6  & 7.3   & 5.4 & 5.0 \nl
H$\alpha$           & 47.6 & 76.1  & 57 & 52  \nl
[N {\sc ii}]$\lambda$6583  & 13.5 & 21.6  & 16  & 15  \nl
[S {\sc ii}]$\lambda$6717  & 8.0 & 12.6   & 9.6 & 8.9 \nl
[S {\sc ii}]$\lambda$6731  & 5.9 & 9.3    & 7.1 & 6.6 \nl
\tableline
\enddata
\tablenotetext{a}{Observed emission-line flux corrected for the Galactic
                  extinction (see text)}
\tablenotetext{b}{Reddening-corrected emission-line flux (see text)}
\tablenotetext{c}{Observed equivalent width}
\tablenotetext{d}{Rest frame equivalent width}

\end{deluxetable}


\newpage

\begin{deluxetable}{lcccc}
\tablecaption{Positions and photometric properties of the galaxies}
\tablehead{
\colhead{Object ID} &
\colhead{RA(J2000)} &
\colhead{DEC(J2000)} &
\colhead{$m_R$\tablenotemark{a}} &
\colhead{$M_R$\tablenotemark{a}} \nl
\colhead{} &
\colhead{($^h ~ ^m ~ ^s$)} &
\colhead{($^\circ ~ ^\prime ~ ^{\prime\prime}$)} &
\colhead{(mag)} &
\colhead{(mag)} \nl
}
\startdata
I   & 15 34 54.3 & 23 27 59 & 24.5 & $-$18.2 \nl
II  & 15 34 54.5 & 23 27 52 & 24.3 & $-$18.4 \nl
III & 15 34 54.8 & 23 27 44 & 24.7 & $-$18.0 \nl
\enddata
\tablenotetext{a}{Corrected for the Galactic extinction (see text).}
\end{deluxetable}


\newpage

\begin{deluxetable}{lcccccc}
\tablecaption{Emission-line fluxes of the galaxies at $z \simeq 0.5$}
\tablehead{
\colhead{Line} &
\colhead{$\lambda_{\rm rest}$} &
\colhead{$\lambda_{\rm obs}$} &
\colhead{$z$}                 &
\colhead{$F$\tablenotemark{a}} &
\colhead{$EW_{\rm obs}$\tablenotemark{b}} &
\colhead{$EW_{\rm rest}$\tablenotemark{c}} \nl
\colhead{} &
\colhead{(\AA)} &
\colhead{(\AA)}  &
\colhead{}                        &
\colhead{($10^{-17}$ erg cm$^{-2}$ s$^{-1}$)} &
\colhead{(\AA)} &
\colhead{(\AA)} \nl
}
\startdata
\multicolumn{7}{l}{Galaxy I} \nl
\tableline
Mg {\sc ii}    & 2798   & 4275.9 & 0.5282 & 2.7 & 28  & 18 \nl
[O {\sc ii}]   & 3727.4 & 5695.9 & 0.5281 & 5.9 & 220 & 144 \nl
[Ne {\sc iii}] & 3868.8 & 5912.7 & 0.5283 & 0.9 & 27  & 18 \nl
H$\beta$       & 4861.3 & 7428.1 & 0.5280 & 1.8 & 57  & 37 \nl
[O {\sc iii}]  & 4958.9 & 7577.0 & 0.5280 & 3.1 & 162 & 106 \nl
[O {\sc iii}]  & 5006.9 & 7650.5 & 0.5280 & 8.1 & 208 & 136 \nl
\tableline
 & & & $<z>=0.5281\pm 0.0001$ & & & \nl
\tableline
\multicolumn{7}{l}{Galaxy II} \nl
\tableline
[O {\sc ii}]   & 3727.4 & 5701.9 & 0.5297 & 2.8 & 37  & 24 \nl
[O {\sc iii}]  & 5006.9 & 7652.0 & 0.5283 & 1.2 & 27  & 18 \nl
\tableline
 & & & $<z>=0.5290\pm 0.0007$ & & & \nl
\tableline
\multicolumn{7}{l}{Galaxy III} \nl
\tableline
[O {\sc ii}]   & 3727.4 & 5674.8 & 0.5225 & 2.8 &  56  & 37 \nl
\tableline
\enddata
\tablenotetext{a}{Observed emission-line flux corrected for the Galactic
                  extinction (see text)}
\tablenotetext{b}{Observed equivalent width.
                  Measurement errors are $\pm$50\% for all the lines.}
\tablenotetext{c}{Rest-frame equivalent width}
\end{deluxetable}


\newpage

\begin{deluxetable}{lcc}
\tablecaption{Star formation rates of the galaxies}
\tablehead{
\colhead{Object ID} &
\colhead{$L$([O {\sc ii}]) ($10^{40}$ erg s$^{-1}$)} &
\colhead{SFR ($M_\odot$ y$^{-1}$)} \nl
}
\startdata
I   & 8.8 & 4.4 \nl
II  & 4.1 & 2.1 \nl
III & 4.0 & 2.0 \nl
\enddata
\end{deluxetable}


\begin{references}
\reference{1}{Arp, H. 1987, ``Quasars, Redshifts and Controversies''
              (Berkeley, Interstellar Media)}
\reference{1}{Baldwin, J. A., Phillips, M. M., \& Terlevich, R. 1981,
              PASP, 93, 5}
\reference{1}{Burstein, D., \& Heiles, C. 1984, \apjs, 54, 33}
\reference{1}{Cowie, L. L., Hu, E. M., \& Songaila, A. 1995, Nature, 377, 603}
\reference{1}{Cowie, L. L., Songaila, A., \& Hu, E. M. 1991, Nature, 354, 460}
\reference{1}{Coziol, R., Ribeiro, A. L. B., de Carvalho, R. R., \&
              Capelato, H. V. 1998, ApJ, 493, 563}
\reference{1}{Dopita, M. A.,  \& Sutherland, R. S.  1996, \apjs, 102, 161}
\reference{1}{Edmunds, M. G., \& Pagel, B. E. J. 1984, MNRAS, 211, 507}
\reference{1}{Ellis, R. S. 1997, ARA \& A, 35, 389}
\reference{1}{Emerson, J. P., Clegg, P. E., Gee, G., Griffin, M. J.,
              Cunningham, C. T., Brown, L. M. J., Robson, E. I., \& Longmore,
              A. J. 1984, Nature, 311, 237}
\reference{1}{Genzel, R., et al. 1998, \apj, 498, 579}
\reference{1}{Guzm\'an, R., Koo, D. C., Faber, S. M., Illingworth, G. D.,
              Takamiya, M., Kron, R. G., \& Barshady, M. A. 1996, ApJ, 460, L5}
\reference{1}{Heckman, T. M., Armus, L., \& Miley, G. K. 1987, \aj, 93, 276}
\reference{1}{Heckman, T. M., Armus, L., \& Miley, G. K. 1990, \apjs, 74, 833}
\reference{1}{Heckman, T. M., Dahlem, M., Eales, S. A., Fabbiano, G., \&
              Weaver, K. 1996, \apj, 457, 616}
\reference{1}{Hibbard, J. E., \& Yun, M. S. 1999, in preparation}
\reference{1}{Hibbard, J. E., \& Yun, M. S. 1996, in ``Cold Gas at High
              Redshift'', edited by~M.~Bremer, H.~Rottgering, ~P.~van der Werf,
              and C. L.~Carilli (Kluwer, Dordrecht), p.~47}
\reference{1}{Hibbard, J. E. \& van Gorkom, J. H. 1996, AJ, 111, 655}
\reference{1}{Hickson, P., Mendes de Oliveira, C., Huchra, J. P., \&
              Palumbo, G. G. C. 1992, ApJ, 399, 353}
\reference{1}{Jarvis, J. F., \& Tyson, J. A. 1981, AJ, 86, 476}
\reference{1}{Kennicutt, R. C. Jr. 1992, ApJ, 388, 310}
\reference{1}{Kinney, A. L., Bohlin, R. C., Calzetti, D., Panagia, N., \& Wyse, R. F. 1993, \apjs, 85, 5}
\reference{1}{Koo, D. C., Guzm\'an, R., Faber, S. M., Illingworth, G. D.,
              Barshady, M. A., Kron, R. G., \& Takamiya, M. 1995, ApJ, 440, L49}
\reference{1}{Koo, D. C., \& Kron, R. 1992, ARA \& A, 30, 613}
\reference{1}{Landolt, A. U. 1983, AJ, 88, 439}
\reference{1}{Meyer, J. -P. 1985, ApJS, 57, 173}
\reference{1}{Oke, J. B., Cohen, J. G., Carr, M., Cromer, J.,
Dingizian, A., Harris, F. H., Labrecque, S., Lucinio, R., Schall, W.,
Epps H., \& Miller, J. 1995, \pasp, 107, 375}
\reference{1}{Osterbrock, D. E. 1989, Astrophysics of Gaseous Nebulae and Active 
              Galactic Nuclei, (University Science Book)}
\reference{1}{Postman, M., \& Lauer, T. 1995, ApJ, 440, 28}
\reference{1}{Sanders, D. B., \& Mirabel, I. F. 1996, ARA \& A, 34, 749}
\reference{1}{Sanders, D. B., Soifer, B. T., Elias, J. H., Madore, B. F.,
              Matthews, K.,  Neugebauer, G., \& Scoville, N. Z.
              1988, ApJ, 325, 74}
\reference{1}{Seaton, M. J. 1979, \mnras, 187, 73p}
\reference{1}{Soifer, B. T., Helou, G., Lonsdale, C. J., Neugebauer, G.,
              Hacking, P.,  Houck, J. R., Low, F. J., Rice, W., \& Rowan-Robinson,
              M. 1984, ApJ, 283, L1}
\reference{1}{Shaya, E., Dowling, D. M., \& Currie, D. G. 1994, AJ, 107, 1675}
\reference{1}{Storchi-Bergmann, T., Kinney, A. L., \& Challis, P. 1995,
              ApJS, 98, 103}
\reference{1}{Taniguchi, Y., \& Ohyama, Y. 1998, ApJ, 508, L13}
\reference{1}{Taniguchi, Y., Trentham, N., \& Shioya, Y. 1998, ApJ, 504, L79}
\reference{1}{Vacca, W. D., \& Conti, P. S. 1992, ApJ, 401, 543}
\reference{1}{Vacca, W. D., Garmany, C. D., \& Shull, J. M. 1996, ApJ, 460, 914}
\reference{1}{Veilleux S., \& Osterbrock, D. E., 1987, ApJS, 63, 295}
\reference{1}{Yun, M. S., Hibbard, J. E., \& Scoville, N. Z. 1999, in
              preparation}
\end{references}
\end{document}